%% file: main-acmtemplate.tex
\documentclass[9pt, sigconf]{acmart}

\usepackage{url}
\usepackage{multirow}
\usepackage{subfigure}
\usepackage{epsfig}
\usepackage{graphicx}
\usepackage{amsmath}
\usepackage{pifont}
\usepackage{color}
\usepackage{setspace}
\usepackage{hhline}   
\usepackage{algorithm}
\usepackage{algpseudocode}
\usepackage{flushend}
\usepackage{url}
\usepackage[normalem]{ulem}
\usepackage{graphics}
\usepackage{tikz}
\usepackage{array}
\usepackage{animate}
\usepackage{titlesec}
\usepackage{balance}
\usepackage{enumitem}
\usepackage{booktabs}

\newcommand{\ignore}[1]{}
\newcommand{\redHL}[1]{\textcolor{red}{#1}}
\newcommand{\blueHL}[1]{{\textcolor{blue}{#1}}}

\titlespacing*{\section}{0pt}{3pt}{3pt}
\titlespacing*{\subsection}{0pt}{3pt}{2pt}
\titlespacing*{\subsubsection}{0pt}{3pt}{2pt}

\DeclareMathOperator*{\minimize}{minimize}

\copyrightyear{2024}
\acmYear{2024}
\setcopyright{acmlicensed}\acmConference[MLCAD '24]{2024 ACM/IEEE International Symposium on Machine Learning for CAD}{September 9--11, 2024}{Salt Lake City, UT, USA}
\acmBooktitle{2024 ACM/IEEE International Symposium on Machine Learning for CAD (MLCAD '24), September 9--11, 2024, Salt Lake City, UT, USA}
\acmDOI{10.1145/3670474.3685945}
\acmISBN{979-8-4007-0699-8/24/09}

\begin{document}

\title[IR-Aware ECO Timing Optimization Using RL]
{IR-Aware ECO Timing Optimization
Using Reinforcement Learning}

\author{Wenjing Jiang}
\affiliation{%
  \institution{University of Minnesota}
  \city{Minneapolis}
  \state{MN}
  \country{USA}
}

\author{Vidya A. Chhabria} 
\affiliation{%
  \institution{Arizona State University}
  \city{Tempe}
  \state{AZ}
  \country{USA}
}

\author{Sachin S. Sapatnekar}
\affiliation{%
  \institution{University of Minnesota}
  \city{Minneapolis}
  \state{MN}
  \country{USA}
}

\begin{abstract}
\noindent
Engineering change orders (ECOs) in late stages make minimal design fixes to recover from timing shifts due to excessive IR drops. This paper integrates IR-drop-aware timing analysis and ECO timing optimization using reinforcement learning (RL). The method operates after physical design and power grid synthesis, and rectifies IR-drop-induced timing degradation through gate sizing. It incorporates the Lagrangian relaxation (LR) technique into a novel RL framework, which trains a relational graph convolutional network (R-GCN) agent to sequentially size gates to fix timing violations. The R-GCN agent outperforms a classical LR-only algorithm: in an open 45nm technology, it (a)~moves the Pareto front of the delay-power tradeoff curve to the left 
(b)~saves runtime over the prior approaches by running fast inference using trained models, and (c) reduces the perturbation to placement by sizing fewer cells. The RL model is transferable across timing specifications and to unseen designs with fine tuning.
\end{abstract}	

\maketitle

\input{sec/1-intro}

\input{sec/2-background-LR}

\input{sec/3-RL-LR-framework}
\input{sec/4-RL-LR-training}

\input{sec/5-results}
\input{sec/6-conclusion}


\bibliographystyle{misc/ACM-Reference-Format}
\bibliography{references}

\ignore{\redHL{\em A couple of other things we need to modify in the writeup:
(a) A lot of this writeup currently (one paragraph in the intro and almost every other paragraph in Section III) is in "defense" mode, i.e., comparing with RL-Sizer. We do not need to go into as many details comparing the smallest of details as the problem statement is itself different now. {\bf (b) This paper was written for DATE submission in Sept 2022. However, between Sept 22 and Dec 22, a lot of the code was updated. Some of the algorithms have changed. For example, there is clock decay aspect which I think is crucial for enabling transferability at different startpoints. However, there is no mention of it in the training flow. Clock update is just one examples, there maybe more changes that have gone into the codee on the algorithm side but not gone into this paper here. Wenjing, please check if the writeup reflects the different lines in the code during training. (c) I am not sure RL-Sizer is the only prior work we should be looking at. First we need to check 
 if any prior art where RL has been used for ECO that we missed. We need some prior work, in IR-aware timing and ECO and not just sizing as more of the references are actually sizing papers.  (c) The results section needs to present the three flows at the start. inference (across designs and startpoints), training flow, and few-shot training flows, then the actual results. I can think we remove the sentence on ASAP7 libs.  Fig. 6 and Fig. 7 can be merged into one figure.} 
}
}

\end{document}

%% file: sec/1-intro.tex
\section{Introduction} 
\label{sec:intro}

\noindent
Tools and flows for power integrity~\cite{OpeNPDN} aim to restrict IR drops below a specified limit, and timing optimization uses gate delay models at this worst-case voltage corner.  However, due to limited wiring resources, in late stages of design, the power grid may not meet the IR limit exactly, resulting in increased gate delays and timing failures. By then, the layout is near-final and large perturbations will impede timing closure; only incremental engineering change order (ECO) optimizations, with minimal placement perturbation, are allowable to resolve timing. We address the ECO timing optimization problem to fix timing violations through gate sizing.
 
Gate sizing for early design stages has been well studied, but there is little work on ECO for IR-induced timing failures. Gate sizing selects a size for every netlist instance from a set of choices in the standard cell library, each with different delay/area/power, and is NP-hard~\cite{tilos}. Early approaches used simple convex models in a space of continuous gate sizes, using sensitivity methods~\cite{sensitivity-abk}, convex programming~\cite{convex}, and Lagrangian relaxation (LR)~\cite{chu2}.
The modern version uses more complex nonconvex delay models and discrete gate sizes~\cite{chu1}.
ECO-based optimizations in~\cite{Chang13,Lee14} do not address the interaction between IR drop and gate sizing.
In~\cite{Lin20}, ECOs for IR drop are resolved by moving cells; results show large placement perturbations. Our ECO solution performs low-perturbation sizing to meet timing and is based on reinforcement learning (RL).  

RL has been applied to several chip design problems~\cite{gcn-rl,survey-rl}. Prior methods have tackled the related gate sizing problem at early stages of design, without supply voltage awareness~\cite{rl-sizer,lr-gnn-sizer}. In~\cite{rl-sizer}, RL is applied in a black-box framework, and its solution is not suitable for ECO optimization; in~\cite{lr-gnn-sizer}, imitation learning is used rather than RL, merely accelerating LR methods to explore delay-power Pareto tradeoffs rather than fully harnessing RL.

Prior RL methods for gate sizing cannot be trivially extended to the ECO problem, but it is useful to examine their limitations.  \textit{First}, they incur \textbf{forbidding runtimes} due to the large action and search spaces for optimization.  \textit{Second}, they focus on a \textbf{single objective} (e.g., minimizing TNS in~\cite{rl-sizer}), rather than the full multi-objective constrained optimization problem. Using penalty- and weight-based techniques that combine power, area, and timing into a single loss function requires significant parameter tuning per design~\cite{mo-rl} and is not viable. \textit{Third}, they use RL as a \textbf{``black box'' optimizer} rather than weaving in prior advances in gate sizing to build a solution that combines the best of traditional and ML-based solvers.

We overcome the first issue through the very nature of our ECO problem formulation: the requirement for minimal perturbation to the existing solution naturally results in action and search spaces of manageable size. To address the second and third issues, we eschew the black-box approach, and instead, leverage domain- and problem-specific knowledge, \textit{using LR to drive RL} by coupling novel RL techniques with essential ideas from prior LR-based approaches.  The use of Lagrange multipliers (LMs) from LR also provides a \textit{natural way to solve constrained optimization problems}.  The use of problem-specific insights also addresses the first issue by improving the efficiency of RL, which is well known to be sample-inefficient~\cite{rlblogpost}. 

We propose \textbf{RL-LR-Sizer}, coupling deep RL with traditional LR-based techniques for ECO gate sizing.  We iteratively answer one of the following questions in each RL iteration: (i)~{\it Order}: ``Given a circuit netlist, which gate to operate on (upsize or downsize)?", and (ii)~{\it Choice}: ``Given a gate and its local neighborhood, what size to select?" 
RL-LR-Sizer uses a relational graph convolutional neural network (R-GCN) as an agent to solve the order and choice problems.
The agent is trained using deep Q learning~\cite{dqn}, interacting with the environment
to maximize a reward. 
We designate a reward function that converts the constrained optimization problem into an unconstrained problem using LMs.
Our contributions are:\\
(1)~This is the first work to address ECO sizing for library-based NLDM delay models under an RL-based framework.\\ 
(2)~We couple RL with LR-based techniques to train an R-GCN agent to determine both \textit{order} and \textit{choice}. This naturally enables \textit{multi-objective optimization}, presenting the first RL formulation for constrained gate sizing. We also leverage problem-specific knowledge in gate sizing to help RL find better solutions.\\
(3)~Our novel clock update method (Sec.~\ref{sec:training}) during training enhances model quality over a range of timing specifications.\\
(4)~We train the RL-LR sizer and use it for inference for ECO changes at multiple timing specifications across multiple designs. We show (a)~a full training flow per design; (b)~zero-shot inference flow using a trained model on an unseen timing specification or unseen design; and (c)~fine tuning flow on a trained model.

%% file: sec/2-background-LR.tex
\section{Problem formulation}
\label{sec:background-lr}

\noindent
The ECO problem is encountered late in the design cycle, after place-and-route and power grid design. 
Larger-than-expected IR drops result in increased gate delays, and the circuit fails timing. The ECO step resizes devices to bring the circuit back to timing specifications.  This involves both device upsizing (to improve drive strength) and downsizing (to reduce the load offered to the previous stage). In principle, upsizing/downsizing could change the current load and alter the voltage drop, but empirically, the change in current load is small, resulting in supply voltage changes (0.001\% of a 1.1V supply).

Formally, the objective of IR-aware ECO timing is to minimize the total power of the design while satisfying performance and electrical constraints after considering post-PD IR drop. 
The constrained optimization problem is formulated as:
\begin{align}
    \;&\; \textstyle \sum_{i\in\mathcal{I}} Power_{c_i} \;\; \label{eq:primal-problem} \\
    \text{subject to}  \;&\; -\mathrm{slack}_i(V_{dd,i},GND_i) \le 0 \; \; \forall i \in \mathcal{I} \nonumber \\
    \;&\; c_i \in \mathcal{C}_i \;\; \forall i \in \mbox{  set of instances  }\mathcal{I} \nonumber
\end{align}
where $\mathcal{C}_i$ is the set of choices for instance $i$ in the library; $c_i \in \mathcal{C}_i$ is the library cell for instance $i$; $Power_{c_i}$ is the power consumption of instance $i$, computed using a vectorless approach and including internal, switching, and leakage power; and slack$_i$ is the slack at instance $i$ as a function of post-PD rail voltages, $V_{dd,i}$ and $GND_i$. 

Problem~\eqref{eq:primal-problem} may be solved using penalty functions~\cite{BV09}, minimizing a weighted sum of the objective and the constraint violations.

%% file: sec/3-RL-LR-framework.tex
\section{RL-LR ECO timing framework}
\label{sec:rl-lr}
\noindent
The RL-LR ECO timing framework solves~\eqref{eq:primal-problem} by: \\
\textbf{(a)}~\textit{Representing} the netlist as a graph with node-level features;\\
\textbf{(b)}~\textit{Mapping} the sizing problem to an RL-solvable control problem; \\
\textbf{(c)}~\textit{Developing} a deep Q-network (DQN) framework, coupled with an LM update strategy, for training the R-GCN model (RL agent).

\subsection{The circuit netlist as an annotated graph}
\label{sec:netlist_graph}

\noindent
The circuit netlist is represented as a directed relational graph $G=(\mathcal{V}, \mathcal{E},\mathcal{R})$ where $\mathcal{V}$ is a set of all nodes representing the instance in the design, $\mathcal{E}$ is a set of all edges representing the nets in the design, and $\mathcal{R}$ is a set of all possible relation types that represent the input or output relation of each edge to the node in the graph. We convert the hyperedges in the circuit to a star representation~\cite{fastplace} where each wire from instance $i$ to instance $j$ is represented by a edge $(v_i, r, v_j) \in\mathcal{E}$ with relation $r$. The graph representation of the netlist allows us to solve the gate sizing problem through the use of deep RL algorithms.

The nodes in the graph are annotated by a list of features (Table~\ref{tbl:features}), used by the R-GCN agent that are capable of understanding the relations between nodes to make decisions. In our work, R-GCNs understand edge directions as the gate delay is dependent on identifying the driver and loads. R-GCNs accumulate annotated features from the set of neighbors, $\mathcal{N}_i^r$, of instance $i$ under relation $r \in \mathcal{R}$. From outgoing edges and incoming edges, R-GCN aggregates different features from neighbors. The output load capacitance is only aggregated from outgoing edges, and the input slew is only aggregated from incoming edges. All other features are aggregated from both edges. The slack, slew, load and IR voltage drop features 
help the R-GCN agent make decisions that meet timing constraints.  The instance type feature enables the agent's decisions to minimize power and meet timing.

\begin{table}[t]
\centering
\caption{List of the annotated features on instance $i$ in $G$.}
\label{tbl:features}
\vspace{-4mm}
\noindent
\resizebox{0.9\linewidth}{!}{
\begin{tabular}{|p{0.22\linewidth}|p{0.8\linewidth}|}
\hline
Feature                        & Definition \\ \hline
$\mathrm{slack}_i$               & Slack at the output pin.    \\
$\mathrm{in\_slew}_i$             & Maximum slew at the input pins.  \\
$\mathrm{out\_slew}_i$            & Slew at the output pin. \\
$\mathrm{instance\_type}$          & Size and type of cell.       \\   
$\mathrm{load}_i$              & Capacitive load at the output pin.               \\
$\mathrm{ir\_voltage}_i$              & Power supply voltage at th cell with IR drop considerd.\\
\hline
\end{tabular}
}
\vspace{-4mm}
\end{table}

\begin{figure}[t]
\centering
\includegraphics[width=0.78\linewidth]{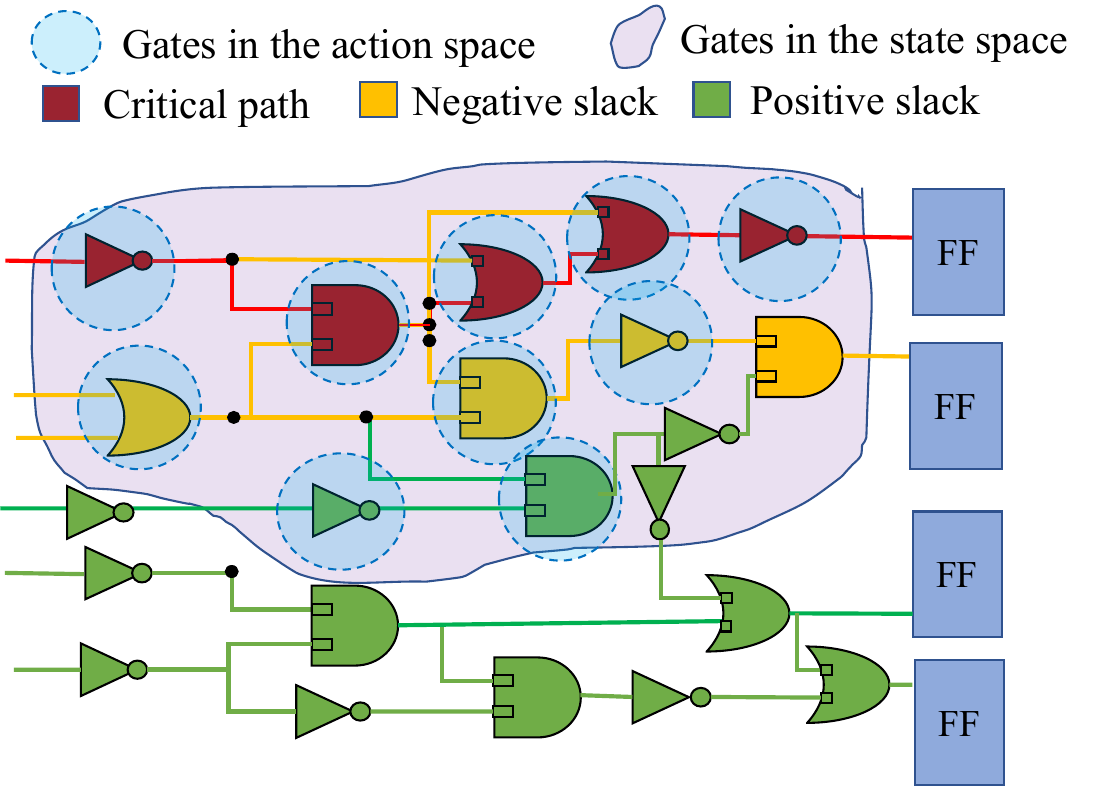}
\vspace{-4mm}
\caption{State and action spaces from the circuit netlist.}
\vspace{-4mm}
\label{fig:state-action}
\end{figure}

\subsection{Mapping IR-aware gate sizing to RL}

\noindent
We map the sizing problem~\eqref{eq:primal-problem} to a Markov Decision Process (MDP), making sequential RL-based decisions. We define the following:

\noindent
{\bf State:} The state $s_t$ at step $t$ is a subgraph of the annotated graph $G$ that contains all instances $v_i$ that have $\mathrm{slack}_i < 0 $, or lie within a two-hop neighborhood\footnote{A two-hop neighborhood is sufficient as the timing impact the choice of a gate size has on the netlist diminishes with the increase in the hops.} of any instance with $\mathrm{slack}_i < 0 $. Unlike~\cite{rl-sizer}, which uses a local embedding of a single independent instance and its three-hop neighborhood as the state, we use \textit{\textbf{all}} netlist instances with negative slack and their neighborhood as the state, letting the RL agent decide which instances in the state to operate on. 

Fig.~\ref{fig:state-action} illustrates the state $s_t$ at time step $t$ (on this toy example, $s_t$ is a large fraction of the circuit, but for a large circuit [des\_perf. 21k gates], $s_t$ contains \textit{only 0.06\% of the gates}). The red instances correspond to gates on the critical path (most negative slack) after considering IR drop impact, the yellow instances correspond to gates that also have negative slack (near-critical paths) and the gates in the purple region are those that belong to the two-hop neighborhood of any instance with $\mathrm{slack}_i < 0 $ (red or yellow instances). This state representation provides the RL agent with a global view of the graph, preventing the optimizer from being stuck within local minima.  In contrast,~\cite{rl-sizer, lr-gnn-sizer, chu1} operate with a single instance (the instance being sized in the current iteration), and its immediate neighborhood as the state. Mimicking these methods would provide an RL agent with little local information, leading to local minima.



\noindent
{\bf Action:} 
We define an action as both the order (which gate is selected) and choice (whether it is upsized or downsized). 
Instead of using all nodes in the state as the action space, we prune the space for faster convergence of RL training by creating an action mask:
\begin{equation}
    \phi(x_a) = 
    \begin{cases} 
        x_a & \text{if $a$ is a valid action} \\ 
        0 & \text{otherwise} \\
    \end{cases}
\end{equation}

The mask constrains the agent to select gates that are on, or within a two-hop neighborhood of any gate on the critical path. The size of the action space, which is based on the {\em critical path} only, is a small fraction of the state space for large circuits.
The mask also prevents invalid actions, e.g., upsizing a gate at its largest available size. In Fig.~\ref{fig:state-action}, all instances highlighted in dotted circles are within the action space. The action space size is twice the number of candidate gates, as each gate can be either upsized or downsized. 

Based on the state and action spaces, we set up the gate sizing RL formulation.
However, the action space is large, creating challenges for the already sample-inefficient RL algorithm~\cite{rlblogpost}.  \textit{We limit the size of the action space by coupling RL training with problem-specific knowledge from LR-based gate sizing, as described Section~\ref{sec:problem-specific}}.

\noindent
{\bf Reward:} We use \textbf{domain-specific knowledge} to create the reward function: instead of using an arbitarily-weighted penalty function to translate~\eqref{eq:primal-problem} to an unconstrained objective, we leverage prior work on Lagrangian relaxation for gate sizing~\cite{chu1} to create an unconstrained reward function.
Using nonnegative Lagrange multipliers (LMs), $\lambda_i$, for each slack constraint, we minimize:
\begin{equation}
\boldsymbol{L_\lambda} (c,\mathrm{slack}): 
\textstyle
\sum_{i\in\mathcal{I}} Power_{c_i} \\
+ \sum_{i\in\mathcal{I}} L_{slack,i}
\label{eq:objective}
\end{equation}
\begin{equation}
\mbox{where   }
L_{\text{slack},i} =
   \begin{cases} 
        \frac{\lambda_i (- \mathrm{slack}_i)}{\beta \times \mathrm{TNS} + \epsilon_0}   & \text{if $\mathrm{TNS} \leq \alpha\times\mathrm{clk}$ } \\ 
        \lambda_i (- \mathrm{slack}_i) & \text{otherwise} \\
    \end{cases}
\end{equation}
and $\boldsymbol{\lambda}$ is the vector of $\lambda_i$s.
Here, $\alpha$ and $\beta $ are tunable constants and $\beta < 0$, $\epsilon_0$ is a small value to prevent divide-by-zero; $\mathrm{clk}$ is the clock period; TNS is the total negative slack; and $c$ [$\mathrm{slack}$] is a set variable for $c_i$ [$\mathrm{slack}_i$]. We formulate $L_{\text{slack},i}$ to provide a higher importance to the slack component of the Langrangian for slacks near zero. 

We will solve the unconstrained Lagrangian relaxation subproblem (LRS) iteratively, updating the LMs $\boldsymbol{\lambda}$, as described in Section~\ref{sec:rl-lr}. We embed the LRS objective into the EL reward function:
\begin{equation}
    R_t = L_{\lambda}^t(c,\text{slack}) -L_{\lambda}^{t+1}(c,\text{slack})
    \label{eq:reward}
\end{equation}
where $L_{\lambda}^t(c,\text{slack})$ and $L_{\lambda}^{t+1}(c,\text{slack})$ is the value of the LRS-based objective function at steps $t$ and $t+1$, respectively, 
due to action $a_t$. Since $a_t$ changes the size of only one gate, with every action, we perform an incremental timing update to evaluate $L_{\lambda}^{t+1}(c,\text{slack})$.

\noindent
{\bf RL agent (R-GCN):} 
GCN-based agents that work on a graph create effective representations of the circuit that encode features into a meaningful embedding through a message passing scheme.

\begin{figure}[t]
\centering
\includegraphics[width=\linewidth]{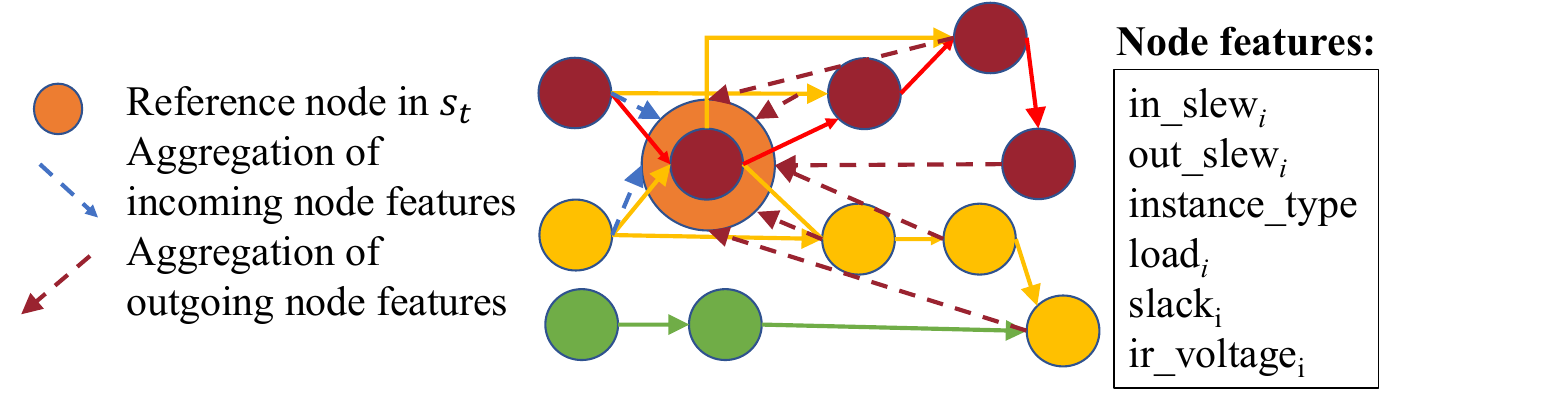}
\caption{R-GCN aggregation from a 3-hop neighborhood.}
\vspace{-4mm}
\label{fig:feature-aggregation}
\end{figure}

The aggregation for the ECO sizing problem is shown in Fig.~\ref{fig:feature-aggregation}, for the reference node highlighted in orange, corresponding to the state in Fig.~\ref{fig:state-action}.  The propagation model for calculating the forward-pass update of a node, $v_i$, in the circuit graph is defined by the following computation performed in layer $l$ of the R-GCN:
\begin{equation}\label{eq:rgcn}
    h_i{(l+1)} = \sigma \left( W_o(l)h_i(l) + \sum_{r\in\mathcal{R}} \sum_{j\in\mathcal{N}_i^r} \frac{1}{d_{i,r}} W_r(l)h_j(l) \right)
\end{equation}
where $\mathcal{N}_i^r$ is the set of neighbors, and  $d_{i,r}$ the normalizing constant, for instance $i$ that have the relation $r$. The relation $r$ is the edge direction in our R-GCN. {\em The use of this R-GCN, rather than a GCN, is critical in capturing the direction of timing flow from input to output of a gate}. In layer $l$, $h_i(l)$ is the hidden state of the node $v_i$; $W_r(l)$ is the weight matrix of the neural network layer $l$ for the relation $r$. The accumulated features are separately and linearly combined for each direction type with the weight matrix $W_r(l)$. The values are normalized by the degree of the node $i$ for the relation $r$, $d_{i,r}$. $W_o(l)$ is the weight matrix for self features; and $\sigma(.)$ is an activation function (we use $\mathrm{ReLU}$). This propagation model is applied to subgraph $s_t$  to generate $h_i(l+1)$ in layer $l+1$.



\begin{figure}[t]
\centering
\includegraphics[width=0.9\linewidth]{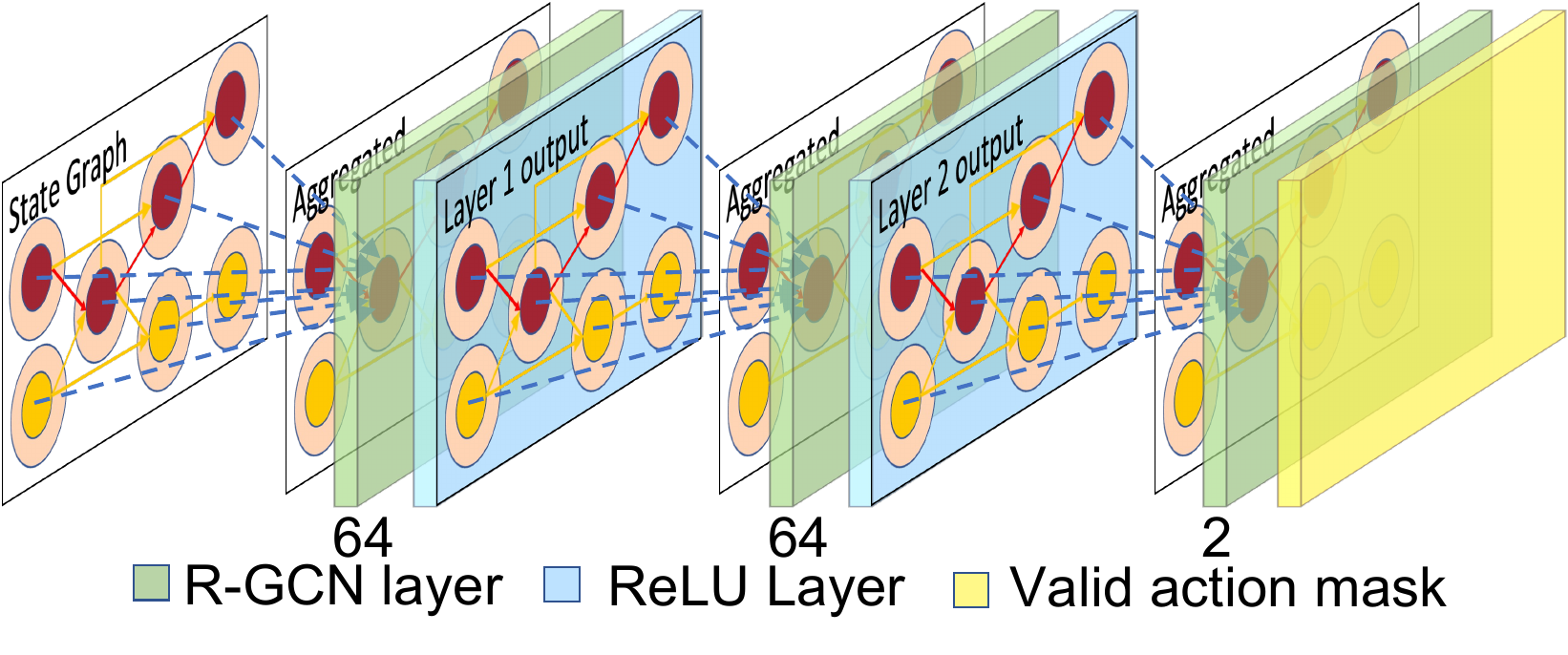}
\vspace{-4mm}
\caption{Structure of the three-layered R-GCN agent.}
\vspace{-4mm}
\label{fig:gcn}
\end{figure}

Fig.~\ref{fig:gcn} shows the structure of the R-GCN agent, with three R-GCN layers, i.e., each node aggregates features from a three-hop neighborhood. The first two layers, each of dimension 64, are followed by ReLU activation functions. The last layer is fed to the action mask to generate a valid action that maximizes the reward.

\noindent
{\bf Environment (Env):} The environment includes components that interact with the agent: in our case, this is the inbuilt timing engine that estimates the reward and updates the graph based on an action.  

\subsection{Application of the R-GCN agent}
\label{sec:rl-lr-appl}

\begin{figure}[t]
\centering
\includegraphics[width=\linewidth]{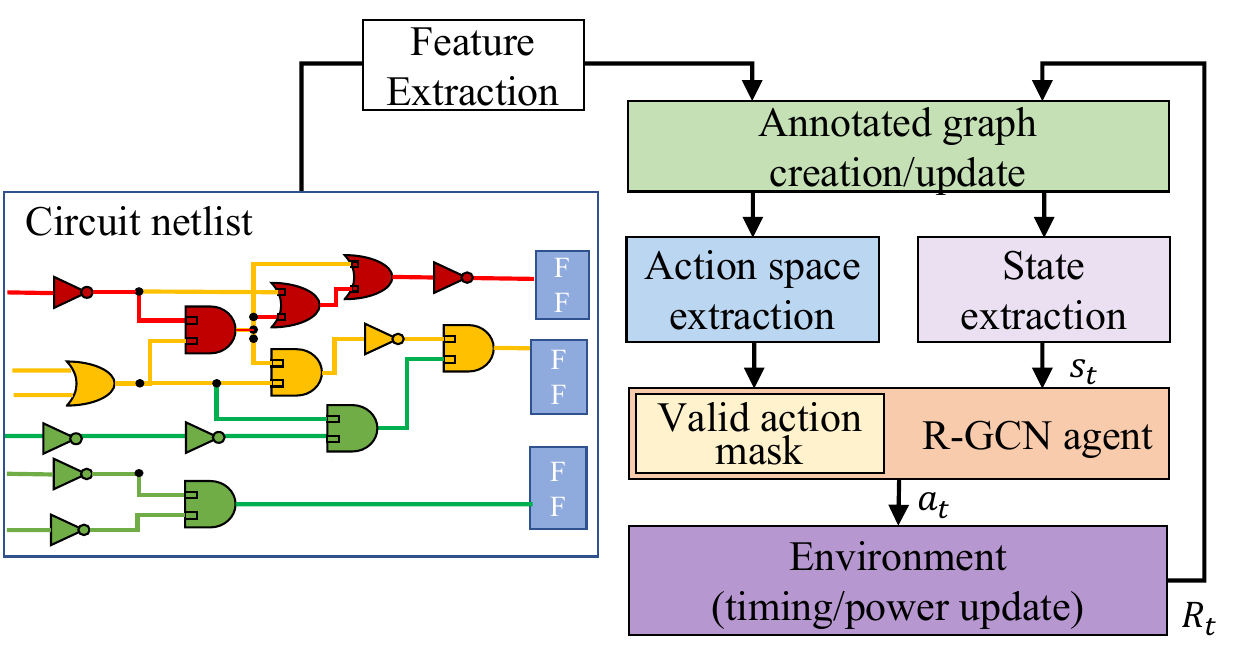}
\vspace{-8mm}
\caption{Overview of the RL-based ECO gate sizing flow.}
\vspace{-4mm}
\label{fig:rl-lr-applied}
\end{figure}

\noindent
We apply our R-GCN agent on a post-P\&R netlist. Fig.~\ref{fig:rl-lr-applied} shows the gate sizing flow and how the R-GCN agent interacts with the environment. Based on the netlist graph $G=(\mathcal{V}, \mathcal{E}, \mathcal{R})$, we extract the node-level features to create $s_t$ and the action mask $\phi$. The R-GCN agent acts on $s_t$ to perform action $a_t$. As a consequence of $a_t$, the environment performs incremental timing analysis, updates the circuit netlist, and computes $R_t$. A new state is created using the updated netlist and the new features, and this is repeated until TNS, WNS, and power converge.

%% file: sec/4-RL-LR-training.tex
\section{RL model training}
\label{sec:training}

\subsection{Core training strategy}

\noindent
We use a deep-Q network (DQN) training algorithm~\cite{dqn} coupled with an LM update strategy~\cite{chu1} to train the R-GCN.  
In the DQN framework, the R-GCN, known as the Q network ($Q(s,a;W)$), is an approximator for the best action-value function, $Q^*(s,a)$, to select actions that maximize expected cumulative reward, defined as:
\begin{equation}
    Q^*(s,a) = \mathbb{E}\left[R+\gamma \max_{a'}Q^*(s',a'|s,a)\right]
\end{equation}
This expression obeys the Bellman equation, which is based on the idea that if the optimal value $Q^*(s', a')$ in the next time step is known for all possible actions $a'$ from state $s'$, the optimal strategy selects action $a'$ to maximize $\mathbb{E}[R + \gamma Q^*(s', a')]$, where $\mathbb{E}(.)$ is the expected value; future rewards are discounted by $\gamma$ per time step $t$. 

\noindent
{\bf DQN training:}
We leverage the DQN training framework described in~\cite{dqn}  which begins by initializing a \textbf{memory replay buffer} of capacity $N$, and random initial R-GCN weights. The training iterates over $M$ \textbf{episodes}, with each episode containing $T$ timesteps.  During each  timestep, an action is taken based on an $\epsilon$-greedy policy strategy where it selects an action from the R-GCN with probability $(1-\epsilon)$ and selects a random action with probability $\epsilon$.  Initially, the training begins with a high value for $\epsilon$, encouraging the agent to explore the environment, and decays every episode to a smaller value exploiting the knowledge the agent has gained. We apply the action mask and select an action from the available valid candidates.

Based on the chosen valid action $a_t$, the environment updates the power and timing incrementally and computes $R_t$. The transition which includes $s_t$, $s_{t+1}$, $a_t$, and $R_t$ is stored in a replay buffer at each step. The buffer is sampled at every step to extract a batch of random samples that train the R-GCN policy network. To make training more stable, the DQN training uses a target network, which keeps a copy of R-GCN weights to estimate the $Q^*(s', a')$ value in the Bellman equation.  The target network is updated every $C$ episodes by copying the weights from the policy network. The R-GCN policy agent is trained by minimizing the loss function:
\begin{equation}
    Loss(W) = \mathbb{E}_{s,a,R}\left[(\mathbb{E}_{s'}[y|s,a] - Q(s,a;W) )^2 \right]
\end{equation}
where $y = \mathbb{E}_{s'}[R+\gamma \max_{a'} \hat{Q}(s',a',\hat{W})|s,a]$ uses the target network $\hat{Q}(s',a',\hat{W})$ to estimate the discounted future rewards.
The target network weights $\hat{W}$ are fixed when optimizing the loss function. 

\noindent
{\bf Env reset:} We use a modified environment reset strategy in comparison to~\cite{dqn}. At the beginning of each episode, instead of resetting the graph to its original state at the beginning of the first episode, we reset the graph to the state which has the least $ \mathbf {L_{\lambda}}(c,slack)$ in the previous episode. This allows for faster convergence of the objective function during training iterations due to guided explorations.   

\subsection{Problem-specific training enhancement}
\label{sec:problem-specific}

\noindent
We incorporate gate-sizing-specific optimizations into the general training framework to enhance the quality of the trained model.

\noindent
{\bf Lagrangian multiplier (LM) update.} The use of LMs to determine the relative weights of the components of the cost function is a crucial problem-specific insight used in this work. A direct application of RL would use fixed user-defined weights for $\lambda_i$, but instead,
we use LR-based update strategies to drive RL exploration. Specifically, we embed the following LM update strategy during DQN training, where we update the values of $\lambda_i$ every $K$ steps within an episode:
$\lambda_i = \lambda_i\times \left(1-\frac{slack_i}{clock\,period}\right)$.
The strategy penalizes large violations more severely than smaller violations. This update strategy helps with guided explorations within the
action space.
When $\mathrm{slack}_i$ is positive or equal to zero, we set the corresponding $\lambda_i$ to zero. 

\noindent
{\bf ECO-based action space reduction.} The ECO problem naturally reduces the sizes of the state and action spaces: given a generally good power grid, it is likely that only some combinational blocks of the sequential circuit in IR-affected regions  will require ECOs. Only gates in these regions will be part of the RL formulation.


\noindent
{\bf Clock constraint update.} The RL framework provides a larger positive reward when the goal is met. A na\"ive RL implementation would provide this reward when the timing specification is met -- but for tight constraints, the RL may find it difficult to reach the specification.  To counter this, recognizing that the delay is progressively reduced using sizing operations, we provide a set of intermediate timing levels as goals that achieve this larger reward, progressively tightening the goal until the delay reaches the specification. 

In the first $M_{decay}$ of $M$ total episodes, the clock constraint decays linearly: at the start of each episode, the clock constraint is reduced by $\frac{(\text{initial\_delay} - \text{target\_delay})}{\mathrm{M}_{decay}}$, where initial\_delay is the delay in post-placement and post-PDN synthesis with IR-drop impact considered and target\_delay is the required delay.
After $\mathrm{M}_{decay}$ episodes, the clock constraint reduces to target\_delay, and is kept unchanged in the remaining training episodes. We set $\mathrm{M}_{decay}=30$. 

To facilitate transferability across multiple designs and start points, we encapsulate away design-specific information and clock information into the state variable. We work with the lambda slack sum, i.e. $\sum_{i\in\mathcal{I}} L_{slack,i}$ in eq.~\eqref{eq:objective}, as the loss function, where the slacks are normalized by clock value set in each episode, instead of absolute delay values, which may vary from circuit to circuit.

\begin{figure}[t]
\centering
\includegraphics[width=1.0\linewidth]{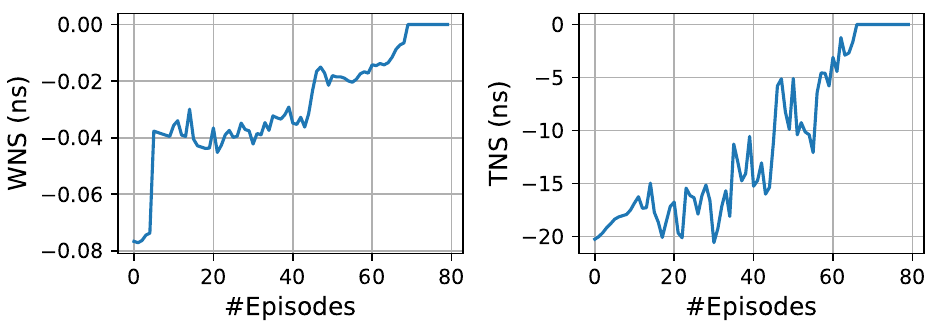}
\vspace{-8mm}
\caption{Results of training, showing the convergence of the WNS and TNS.}
\vspace{-4mm}
\label{fig:tns-converging}
\end{figure}

Fig.~\ref{fig:tns-converging} shows the effectiveness of these approaches in training for circuit wb\_conmax. After $M$ episodes, we see that TNS and WNS have converged. Once trained, the R-GCN agent (target network) is applied to the circuit netlist (see Section~\ref{sec:rl-lr-appl}) to sequentially select actions that solve the constrained optimization problem.

%% file: sec/5-results.tex
\section{Evaluation of RL-LR ECO}
\label{sec:results}

\subsection{Experimental setup}

\begin{figure}[b]
\centering
\vspace{-4mm}
\includegraphics[width=0.9\linewidth]{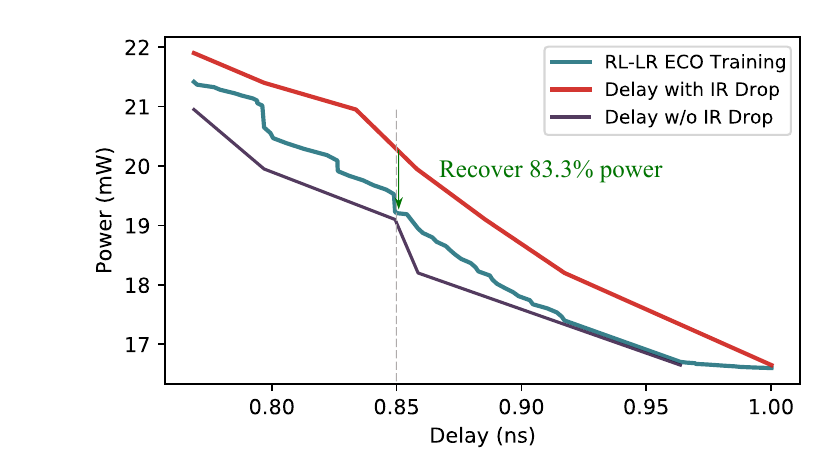}
\vspace{-4mm}
\caption{Power-delay tradeoff curves for wb\_conmax in a 45nm technology, showing a shift to the right due to IR drop, and the results of our RL-LR ECO approach.}
\label{fig:power-delay-tradeoff}
\end{figure}

\noindent
We use Design Compiler to size gate-level netlists under various timing constraints at the supply voltage corner. We use OpenROAD~\cite{openroad} to place the design, synthesize the power delivery network, and analyze the static IR drop. An example delay-power curve for wb\_conmax is shown as ``Delay w/o IR-drop'' in Fig.~\ref{fig:power-delay-tradeoff}. The PDN is built to so that the maximum IR drop is low ($\sim$5mV) or moderate ($\sim$10mV) scenarios where delay degradations can be recovered through ECOs. \ignore{We simulate larger IR drops in the PDN by manually adjusting per-unit RC values in the metal layers to achieve low ($\sim$5mV) and moderate ($\sim$10mV) maximum IR drop across the designs; we do not consider high IR drops that would require major changes rather than ECOs.} To quantify delay shifts due to IR drop, we annotate NLDM entries of each library cell with its delay sensitivity to voltage drop, and compute circuit delays using the OpenSTA timer. \ignore{Due to limitations of the characterized NanGate45 library, we obtain these sensitivities from a similar library in another technology.} Under IR drop, the delay-power curve shifts to the right, from the ``Delay w/o IR drop'' to the ``Delay with IR drop'' curve in Fig.~\ref{fig:power-delay-tradeoff}.

The RL-LR ECO timing (training and application) flow is built using Python libraries, including PyTorch and DGL~\cite{karypis20}.
We use SWIG-based Python enablements for incremental timing analysis, logic gate swap, and pin and cell property from OpenROAD APIs, which integrates easily into Env for reward computation and state transitions. OpenROAD eliminates challenges of slow TCL interfaces between commercial tools and Python environments in~\cite{rl-sizer}. We train the R-GCN agent using the hyperparameters $M=50$; $N=4000$; $T=75$; $\gamma=0.99$; $\alpha= 10$;  $\beta=-0.1$; $C=25$; $K=30$. 

Our testcases are 7 OpenCores benchmarks using an open 45nm technology, with 2k--26k gates.  Table~\ref{tbl:benchmarks} shows the number of instances on paths that fail timing.\ignore{ Table~\ref{tbl:benchmarks} summarizes the total number of gates, the number of gates on failed paths and the corresponding target period for each design.} We implement multiple flows:\\ 
(1)~\textbf{RL-LR ECO Training} (Table~\ref{tbl:within-design-10mV-power}) trains an R-GCN model from scratch, starting from the initial delay, until the target delay is met.\\
(2)~\textbf{RL-LR ECO Inference} uses a model that is trained across a wide range of timing constraints to obtain the full delay-power Pareto curve. This training step is applicable for larger delay reductions than ECO optimizations, but requires more computation than that for ECO training. Given this trained model, we report:\\
(a)~\textbf{Inference across Timing Constraints (Inf-TC)} (Table~\ref{tbl:across-points}) shows transferability for inference across multiple timing specifications,
using a model trained on the same design.\\
(b)~\textbf{Inference across Designs (Inf-D)} (Table~\ref{tbl:across-designs}) demonstrates transferability across designs, running inference for multiple designs using the model trained with four selected designs.\\
(c)~\textbf{Fine Tuning} (Table~\ref{tbl:across-designs}) applies inexpensive design-specific fine tuning to the ``Inference across Designs'' model, 
incrementally updating weights of the pretrained R-GCN model.\\
(3)~\textbf{LR Baseline} (Tables~\ref{tbl:within-design-10mV-power}--\ref{tbl:across-designs}) implements a conventional LR flow~\cite{chu1} against which the RL methods are compared.

\ignore{
\noindent
{\bf LR baseline:} We implement a baseline LR-based logic gate sizing flow~\cite{chu1}. We perform topological traversal of the netlist, and for each gate, we iterate through all the available alternative choices and select the size that minimizes the local cost function defined in~\eqref{eq:lr-subproblem}. We perform several iterations of the traversal until the cost converges and update $\lambda_i$ in the same way described in Section~\ref{sec:training}.  This flow is run for different clock periods to obtain the delay-power tradeoff curve. 

\noindent
\blueHL{{\bf Inference flows:} To compare the quality of the solutions and runtime of trained models to LR baseline and test the transferbility for different start points on the delay-power curve and the transferbility ac, we run the inferences for each design using the model trained across different start points within designs and using the model trained across designs. For each design, we run the inference from several different start points on the delay-power tradeoff curve with IR drop considered.}

\noindent
\blueHL{{\bf Fine tuning flow:} As the quality of the solutions from the inferences using the model trained across the designs is not as good as the solutions from the inferences using the trained across multiple start points within a design, we further tune the model for each test design with much smaller runtime than training from scratch.}
}

All runtimes are reported on an Intel Xeon Silver 4214 CPU @2.2GHz and NVIDIA A100 PCIe 40GB GPU. 

\input{sec/tbl-benchmarks-new}
\input{sec/tbl-within-design-power-overhead-percentage-new}

\subsection{Optimization and delay-power tradeoffs}

As stated earlier, Fig.~\ref{fig:power-delay-tradeoff} shows the power-delay tradeoff for the ideal IR drop (``Delay w/o IR drop'') and the actual IR drop (``Delay with IR drop'') for the circuit wb\_conmax. The third curve shows the result of our RL-LR ECO algorithm, which is seen to recover the delay degradation due to IR drop.
To quantify the effectiveness of ECO, we define a recovery metric: for a specific delay $D$, we define the power overhead savings of the RL-LR ECO approach as $\frac{\text{Power\_IR}(D) - \text{Power\_RL-LR}(D)}{\text{Power\_IR}(D) - \text{Power\_w/o\_IR}(D)}$, where ``IR,'' ``RL-LR,'' and ``w/o\_IR'' refer to the three curves in the figure. 
For example, at $D=0.85$ns, 83.3\% of the IR-induced power overhead is recovered.

Table~\ref{tbl:within-design-10mV-power} compares the traditional LR baseline, RL-LR ECO Training, and Inf-TC for the seven benchmark designs under moderate IR drops, for a specified target delay. The inference model is trained just once and has a training time of 40--70 minutes. The training cost can be amortized over multiple ECO explorations during late-stage design, e.g., perturbations to routing, PDN design, or placement; each translates to altered slack, and is covered by our formulation.

The table shows the initial delay of the circuit, i.e., the delay under the ideal IR drop; the total power and runtimes for the three approaches; and the number of cell changes using the LR and RL-LR ECO training methods. At each target delay, the RL-LR ECO training saves as much or more power than the LR baseline flow, with 29.1\% average and 48.8\% maximum savings.
The inference approach performs slightly worse than RL-LR ECO training in six testcases, but better in one testcase because the solution (a sequence of sizing) from RL-LR ECO training flow mixes a few random actions for exploration purposes during training. All inferences provide solutions of similar quality to LR baseline, with some improvements.

Table~\ref{tbl:within-design-10mV-power} also reports the number of upsized/downsized cells: downsized cells stay in place and do not require further layout changes, but upsized cells require placement perturbations to remove overlaps. Both Inference and RL-LR ECO training upsize fewer cells than LR for large designs, thus easing timing closure.

As is typical of RL-based approaches~\cite{rl-sizer}, RL-LR ECO training has very high runtime as it includes R-GCN training and a reward computation (STA update) during each training step. The Inf-TC model, with offline training, shows an average of 44.5\% runtime reduction over LR. After one-time training, the trained model can be applied to any operating period on the delay-power curve.

\input{sec/tbl-across-start-points-power-overhead-percentage-new}

To demonstrate that the model is transferable across multiple start points within a design, Table~\ref{tbl:across-points} presents results for five start points for wb\_conmax. The result for the Inference flow is similar to the  RL-LR training from scratch, and better than the LR baseline.

\begin{figure}
\centering
\includegraphics[width=0.95\columnwidth]{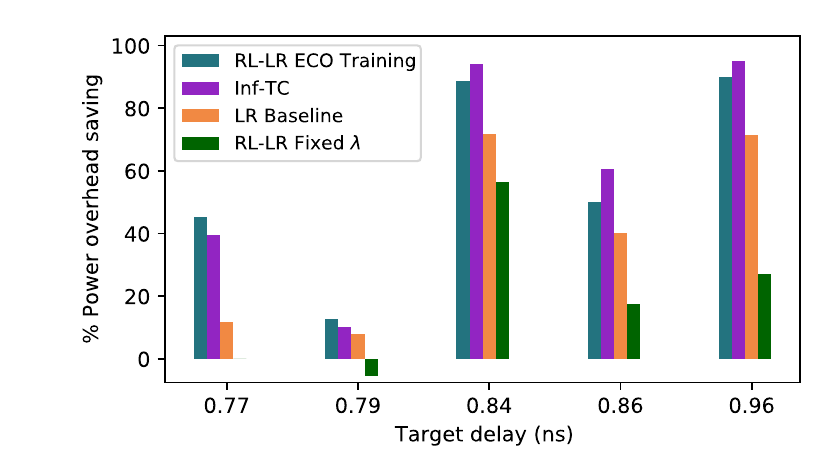}
\vspace{-4mm}
\caption{Power overhead savings for wb\_conmax.}
\label{fig:importance}
\vspace{-5mm}
\end{figure}


\subsection{Importance of LR-based weight adaptation}
\label{sec:importance-LR}

\noindent
We highlight crucial aspects of the RL-LR ECO optimizer: (i)~guided explorations through LR, and (ii)~solving both the gate order and choice (gate size selection) problems, as defined in Section~\ref{sec:intro}.

Fig.~\ref{fig:importance} shows power overhead savings for wb\_conmax, at different target delays, for several flows: our trained RL-LR ECO approach, the classical LR baseline, Inf-TC, and RL-LR with fixed weights, $\lambda$.
Across all target delays, our RL-LR ECO optimizer and Inf-TC methods show better savings than the classical LR solution.

To understand the importance of using LMs to determine the relative weights of the objective function components, dynamically changed during the RL optimization, we consider the use of fixed $\lambda_i$ penalties for the slack term in~\eqref{eq:objective} throughout the optimization (``RL-LR Fixed $\lambda$''). This corresponds to a direct application of RL without problem-specific insights used in our RL-LR method. The fixed-$\lambda$ overhead has negative savings at 0.79ns and cannot meet the tight target specification of 0.77ns.
Thus, fixed user-defined $\lambda_i$ weights are suboptimal as they predetermine the effort that the optimizer makes towards meeting each term in the objective function.
In contrast, RL-LR ECO training and Inference flows use an R-GCN agent that \textit{dynamically} updates the $\lambda_i$ values; both outperform the fixed-$\lambda$ case. The success of our RL-LR methods is achieved by running an RL-based solution of the constrained optimization problem, with automated tuning of the $\lambda_i$s that weight the constraint functions relative to the objective.
\input{sec/tbl-transferbility-power-overhead-percentage}

\subsection{Transferability across designs}
\noindent
To train over a transferable model across multiple designs, we select four designs highlighted in blue in Table~\ref{tbl:across-designs} and perform offline RL training. The table shows that in thirteen testcases, inference (Inference across Designs) using the trained model can achieve the target delay; ``--'' represents an unachievable target delay. 
The model works for all testcases under low ($\le$5mV) IR drop, and all but one in the moderate ($\le$10mV) IR drop regime, as the timing degradation for the former is smaller and easier to fix. We do not consider high IR drops that would require major changes rather than ECOs.

Since training across designs builds a ``common'' model for the four designs, the zero-shot model may fail to meet the target delay of unseen testcases (circuits that are not blue).
If so, we apply inexpensive fine-tuning on the trained model individually for each testcase, and can then achieve the target timing in all cases. We almost always have lower power cost than the LR baseline, and much lower runtime than training from scratch. Designs that successfully meet their target periods do not need fine tuning, as shown by ``NA.'' Our runtime improvement is 55.7\% over LR on average. \textit{In 13 of 14 testcases, fine-tuning is not needed, and our runtime is better than that for LR} (one testcase is tied). For wb\_conmax under moderate IR drop, fine tuning shows power improvement. The runtime column here shows the sum of Inf-D and fine-tuning runtimes.



%% file: sec/tbl-benchmarks-new.tex
\begin{table}[t]
\centering
\caption{IR-induced Timing Failures}
\vspace{-4mm}
\label{tbl:benchmarks}
\hspace*{-3mm}
\resizebox{1.1\columnwidth}{!}{%
\begin{tabular}{|l||c|c|c|c|c|c|c||}
\hline
\textbf{Design} & \rotatebox[origin=c]{45}{des\_area} 
                         & \rotatebox[origin=c]{45}{wb\_dma} 
                                & \rotatebox[origin=c]{45}{pid\_controller} 
                                        & \rotatebox[origin=c]{45}{aes\_cipher\_top}
                                              & \rotatebox[origin=c]{45}{des\_perf}  
                                                    & \rotatebox[origin=c]{45}{pci\_bridge32}
                                                            & \rotatebox[origin=c]{45}{wb\_conmax} 
                                                                                    \\ \hline \hline
\textbf{Target} & \multirow{2}{*}{0. 56}  
                         & \multirow{2}{*}{0.47} 
                                & \multirow{2}{*}{0.75} 
                                        & \multirow{2}{*}{0.88} 
                                              & \multirow{2}{*}{0.59} 
                                                     & \multirow{2}{*}{0.76} 
                                                            & \multirow{2}{*}{0.80} \\
\textbf{period (ns)}
                &        &      &      &      &      &      &      \\ \hline
\textbf{\#Instances on failed}& \multirow{2}{*}{25}
                        & \multirow{2}{*}{36}
                                & \multirow{2}{*}{78}
                                        & \multirow{2}{*}{528}
                                              & \multirow{2}{*}{492}
                                                     & \multirow{2}{*}{220}
                                                            &\multirow{2}{*}{318}\\
\textbf{paths (5mV max. IR)}& & & & & & &\\ \hline
\textbf{\#Instances on failed}
                & \multirow{2}{*}{129}
                         & \multirow{2}{*}{226}
                                & \multirow{2}{*}{182}
                                       & \multirow{2}{*}{1482}
                                              & \multirow{2}{*}{2110}
                                                     & \multirow{2}{*}{1142}
                                                            & \multirow{2}{*}{1530} \\ 
\textbf{paths (10mV max. IR)}&        &      &      &      &      &      &      \\ \hline
\end{tabular}
}
\end{table}

%% file: sec/tbl-within-design-power-overhead-percentage-new.tex
\begin{table}[t]
\centering
\caption{ECO results for three flows: LR baseline, inference across timing constraints (Inf-TC), and RL-LR ECO training}
\vspace{-4mm}
\label{tbl:within-design-10mV-power}
\hspace*{-3mm}
\resizebox{1.1\linewidth}{!}{
\begin{tabular}{|l|c|c|c|c|c|c|c|c|c|c|c|} 
\hline
\multirow{2}{*}{Designs}     & \multirow{2}{*}{\begin{tabular}[c]{@{}c@{}}Target\\Delay\\(ns)\end{tabular}} & \multirow{2}{*}{\begin{tabular}[c]{@{}c@{}}Initial\\Delay\\(ns)\end{tabular}} & \multicolumn{3}{c|}{\% Power Overhead Savings}                                                                                                                                                    & \multicolumn{3}{c|}{Runtime(s)}                                                                                                                                                                    & \multicolumn{3}{c|}{\# of Upsize/Downsize}                                                                                                                                                         \\ 
\cline{4-12}
                             &                                                                              &                                                                               & \begin{tabular}[c]{@{}c@{}}LR\end{tabular} & \begin{tabular}[c]{@{}c@{}}Inf-TC\end{tabular} & \begin{tabular}[c]{@{}c@{}}RL-LR\\ECO\\Training\end{tabular} & \begin{tabular}[c]{@{}c@{}}LR\end{tabular} & \begin{tabular}[c]{@{}c@{}}Inf-TC\end{tabular} & \begin{tabular}[c]{@{}c@{}}RL-LR\\ECO\\Training\end{tabular} & \begin{tabular}[c]{@{}c@{}}LR\end{tabular} & \begin{tabular}[c]{@{}c@{}}Inf-TC\end{tabular} & \begin{tabular}[c]{@{}c@{}}RL-LR\\ECO\\Training\end{tabular}  \\ 
\hline
des\_area                    & 0.56                                                                         & 0.590                                                                        & 35.2\%                                              & 37.0\%                                                                      & 48.8\%                                                     & 82                                                   & 23                                                                  & 2651                                                        & 29/0                                                 & 36/6                                                                         & 27/4                                                         \\ 
\hline
wb\_dma                      & 0.47                                                                         & 0.486                                                                        & 16.7\%                                              & 18.9\%                                                                      & 19.7\%                                                     & 47                                                   & 42                                                                  & 2284                                                        & 12/0                                                 & 19/0                                                                         & 18/3                                                         \\ 
\hline
pid\_controller              & 0.75                                                                         & 0.781                                                                        & 32.6\%                                              & 33.3\%                                                                      & 33.7\%                                                     & 58                                                   & 21                                                                  & 2881                                                        & 29/2                                                 & 29/2                                                                         & 20/9                                                         \\ 
\hline
aes\_cipher\_top             & 0.89                                                                         & 0.941                                                                        & 8.6\%                                               & 13.8\%                                                                      & 27.2\%                                                     & 654                                                  & 153                                                                 & 2305                                                        & 386/3                                                & 301/3                                                                        & 291/1                                                        \\ 
\hline
des\_perf                    & 0.60                                                                         & 0.632                                                                        & 8.3\%                                               & 15.6\%                                                                      & 12.8\%                                                     & 576                                                  & 398                                                                 & 2322                                                        & 491/1                                                & 393/1                                                                        & 380/5                                                        \\ 
\hline
pci\_bridge32                & 0.75                                                                         & 0.781                                                                        & 16.0\%                                              & 16.0\%                                                                      & 16.2\%                                                     & 92                                                   & 54                                                                  & 2336                                                        & 24/1                                                 & 22/1                                                                         & 23/1                                                         \\ 
\hline
wb\_conmax                   & 0.77                                                                         & 0.834                                                                        & 11.8\%                                              & 39.5\%                                                                      & 45.3\%                                                     & 290                                                  & 242                                                                 & 3193                                                        & 86/0                                                 & 78/0                                                                         & 73/0                                                         \\ 
\hline
\multicolumn{1}{|c}{Average} & \multicolumn{1}{c}{}                                                         & \multicolumn{1}{c}{}                                                          & \multicolumn{1}{c}{18.4\%}                          & \multicolumn{1}{c}{24.9\%}                                                  & \multicolumn{1}{c}{29.1\%}                                 & \multicolumn{1}{c}{}                                 & \multicolumn{1}{c}{$-44.5$\%}                                                  & \multicolumn{1}{c}{}                                        & \multicolumn{1}{c}{}                                 & \multicolumn{1}{c}{$+3.3$\%}                                                  & $-7.8$\%                                                      \\ 
\hline
\multicolumn{1}{|c}{Maximum} & \multicolumn{1}{c}{}                                                         & \multicolumn{1}{c}{}                                                          & \multicolumn{1}{c}{35.2\%}                          & \multicolumn{1}{c}{39.5\%}                                                  & \multicolumn{1}{c}{48.8\%}                                 & \multicolumn{1}{c}{}                                 & \multicolumn{1}{c}{$-76.6$\%}                                                  & \multicolumn{1}{c}{}                                        & \multicolumn{1}{c}{}                                 & \multicolumn{1}{c}{$-22.0$\%}                                                 & $-31.0$\%                                                     \\
\hline
\end{tabular}
}
\end{table}

%% file: sec/tbl-across-start-points-power-overhead-percentage-new.tex
\begin{table}
\centering
\caption{Results on multiple clock constraints for wb\_conmax}
\vspace{-4mm}
\label{tbl:across-points}
\resizebox{0.9\columnwidth}{!}{
\begin{tabular}{|c|c|c|c|c|c|c} 
\cline{1-6}
\multirow{3}{*}{Designs}     & \multirow{3}{*}{\begin{tabular}[c]{@{}c@{}}Target Delay\\(ns)\end{tabular}} & \multirow{3}{*}{\begin{tabular}[c]{@{}c@{}}Initial Delay\\(ns)\end{tabular}} & \multicolumn{3}{c|}{\% Power Overhead Savings}                                                                                                                                                                                                   &   \\ 
\cline{4-6}
                             &                                                                             &                                                                              & \multirow{2}{*}{\begin{tabular}[c]{@{}c@{}}LR\end{tabular}} & \multirow{2}{*}{\begin{tabular}[c]{@{}c@{}}Inf-TC\end{tabular}} & \multirow{2}{*}{\begin{tabular}[c]{@{}c@{}}RL-LR ECO\\Training\end{tabular}} &   \\
                             &                                                                             &                                                                              &                                                                       &                                                                                               &                                                                          &   \\ 
\cline{1-6}
\multirow{5}{*}{wb\_conmax}  & 0.77                                                                        & 0.834                                                                       & 11.76\%                                                               & 39.5\%                                                                                       & 45.3\%                                                                  &   \\ 
\cline{2-6}
                             & 0.79                                                                        & 0.858                                                                       & 8.0\%                                                                & 10.3\%                                                                                       & 12.8\%                                                                  &   \\ 
\cline{2-6}
                             & 0.84                                                                        & 0.886                                                                       & 71.7\%                                                               & 94.3\%                                                                                       & 88.8\%                                                                  &   \\ 
\cline{2-6}
                             & 0.86                                                                        & 0.917                                                                       & 40.3\%                                                               & 60.5\%                                                                                       & 50.2\%                                                                  &   \\ 
\cline{2-6}
                             & 0.96                                                                        & 1.0002                                                                       & 71.5\%                                                               & 95.2\%                                                                                       & 90.1\%                                                                  &   \\ 
\cline{1-6}
\multicolumn{1}{|c}{Average} & \multicolumn{1}{c}{}                                                        & \multicolumn{1}{c}{}                                                         & \multicolumn{1}{c}{40.7\%}                                           & \multicolumn{1}{c}{60.0\%}                                                                   & 57.5\%                                                                  &   \\ 
\cline{1-6}
\multicolumn{1}{|c}{Maximum} & \multicolumn{1}{c}{}                                                        & \multicolumn{1}{c}{}                                                         & \multicolumn{1}{c}{71.7\%}                                           & \multicolumn{1}{c}{95.2\%}                                                                   & 90.1\%                                                                  &   \\
\cline{1-6}
\end{tabular}
}
\vspace{-5mm}
\end{table}

%% file: sec/tbl-transferbility-power-overhead-percentage.tex
\begin{table}
\centering
\caption{Transferability across designs}
\vspace{-4mm}
\label{tbl:across-designs}
\hspace*{-2mm}
\resizebox{\columnwidth}{!}{
\begin{tabular}{|c|l|c|c|c|c|c|c|c|} 
\hline
\multirow{2}{*}{IR Drop}                                                       & \multirow{2}{*}{Designs}           & \multirow{2}{*}{\begin{tabular}[c]{@{}c@{}}Target\\Delay\\(ns)\end{tabular}} & \multirow{2}{*}{\begin{tabular}[c]{@{}c@{}}Initial\\Delay\\(ns)\end{tabular}} & \multicolumn{3}{c|}{\% Power Overhead Savings}                                                                                                                                  & \multicolumn{2}{c|}{Runtime(s)}                                                                                                                      \\ 
\cline{5-9}
                                                                               &                                    &                                                                              &                                                                               & \begin{tabular}[c]{@{}c@{}}LR\\Baseline\end{tabular} & \begin{tabular}[c]{@{}c@{}}Inf-D\end{tabular} & \begin{tabular}[c]{@{}c@{}}Fine\\Tuning\end{tabular} & \begin{tabular}[c]{@{}c@{}}LR\\Baseline\end{tabular} & \begin{tabular}[c]{@{}c@{}}Inf-D + \\Fine Tuning\end{tabular}  \\ 
\hline
\multirow{7}{*}{\begin{tabular}[c]{@{}c@{}}Low\\($\le$5mV)\end{tabular}}       & \textcolor{blue}{des\_area}        & 0.56                                                                         & 0.568                                                                        & 76.4\%                                              & 78.7\%                                                           & NA                                                   & 55                                                   & 24                                                                                            \\ 
\cline{2-9}
                                                                               & wb\_dma                            & 0.47                                                                         & 0.475                                                                        & 20.6\%                                              & 21.0\%                                                           & NA                                                   & 39                                                   & 12                                                                                            \\ 
\cline{2-9}
                                                                               & \textcolor{blue}{pid\_controller}  & 0.75                                                                         & 0.764                                                                        & 51.6\%                                              & 51.7\%                                                           & NA                                                   & 54                                                   & 20                                                                                            \\ 
\cline{2-9}
                                                                               & \textcolor{blue}{aes\_cipher\_top} & 0.89                                                                         & 0.908                                                                        & 4.3\%                                               & 16.7\%                                                           & NA                                                   & 213                                                  & 121                                                                                           \\ 
\cline{2-9}
                                                                               & des\_perf                          & 0.59                                                                         & 0.609                                                                        & 14.3\%                                              & 18.4\%                                                           & NA                                                   & 514                                                  & 142                                                                                           \\ 
\cline{2-9}
                                                                               & \textcolor{blue}{pci\_bridge32}    & 0.75                                                                         & 0.772                                                                        & 24.2\%                                              & 24.5\%                                                           & NA                                                   & 185                                                  & 108                                                                                           \\ 
\cline{2-9}
                                                                               & wb\_conmax                         & 0.77                                                                         & 0.829                                                                        & 21.6\%                                              & 23.8\%                                                           & NA                                                   & 274                                                  & 120                                                                                           \\ 
\hline
\multirow{7}{*}{\begin{tabular}[c]{@{}c@{}}Moderate\\($\le$10mV)\end{tabular}} & \textcolor{blue}{des\_area}        & 0.56                                                                         & 0.590                                                                        & 35.2\%                                              & 45.7\%                                                           & NA                                                   & 82                                                   & 36                                                                                            \\ 
\cline{2-9}
                                                                               & wb\_dma                            & 0.47                                                                         & 0.486                                                                        & 16.7\%                                              & 19.1\%                                                           & NA                                                   & 47                                                   & 33                                                                                            \\ 
\cline{2-9}
                                                                               & \textcolor{blue}{pid\_controller}  & 0.75                                                                         & 0.781                                                                        & 32.6\%                                              & 33.7\%                                                           & NA                                                   & 58                                                   & 19                                                                                            \\ 
\cline{2-9}
                                                                               & \textcolor{blue}{aes\_cipher\_top} & 0.89                                                                         & 0.941                                                                        & 8.6\%                                               & 29.8\%                                                           & NA                                                   & 654                                                  & 94                                                                                            \\ 
\cline{2-9}
                                                                               & des\_perf                          & 0.60                                                                         & 0.632                                                                        & 8.3\%                                               & 11.3\%                                                           & NA                                                   & 576                                                  & 175                                                                                           \\ 
\cline{2-9}
                                                                               & \textcolor{blue}{pci\_bridge32}    & 0.75                                                                         & 0.781                                                                        & 16.0\%                                              & 16.2\%                                                           & NA                                                   & 92                                                   & 15                                                                                            \\ 
\cline{2-9}
                                                                               & wb\_conmax                         & 0.77                                                                         & 0.834                                                                        & 11.8\%                                              & -                                                                 & 16.5\%                                              & 290                                                  & 37+294=331                                                                                           \\ 
\hline
\multicolumn{2}{|c}{Average}                                                                                        & \multicolumn{1}{c}{}                                                         & \multicolumn{1}{c}{}                                                          & \multicolumn{1}{c}{24.4\%}                          & \multicolumn{1}{c}{30.0\%}                                       & \multicolumn{1}{c}{}                                 & \multicolumn{1}{c}{}                                 & --55.7\%                                                                                       \\ 
\hline
\multicolumn{2}{|c}{Maximum}                                                                                        & \multicolumn{1}{c}{}                                                         & \multicolumn{1}{c}{}                                                          & \multicolumn{1}{c}{76.4\%}                          & \multicolumn{1}{c}{78.7\%}                                       & \multicolumn{1}{c}{}                                 & \multicolumn{1}{c}{}                                 & --85.6\%                                                                                       \\
\hline
\end{tabular}
}
\vspace{-8mm}
\end{table}

%% file: sec/6-conclusion.tex
\section{Conclusion}
\label{sec:conclusion}
\noindent
Our RL-LR ECO method recovers degradations to the power-delay tradeoff curve due to IR drops.  We incorporate problem-specific knowledge into RL-driven gate sizing. Together with other problem-specific methods, our method shifts  the Pareto optimal front of the power-delay tradeoff curve to the left. Our methods range from a full training flow per design to zero-shot inference on unseen specifications, as well as a fine tuning flow on a pretrained model.
We believe that our LR-based weight update strategy is generalizable for solving any constrained optimization problem using RL.